\documentclass[a4paper,10pt]{article}
\usepackage[utf8x]{inputenc}
\usepackage{graphicx}
\usepackage{amssymb,amsmath}

\newcommand{\mhiggs}{M_h}
\newcommand{\oh}{\Omega h^2}
\newcommand{\gev}{\mathrm{GeV}}
\newcommand{\tev}{\mathrm{TeV}}

\newcommand{\ma}{M_{A^0}}
\newcommand{\mh}{M_{H^0}}
\newcommand{\mhc}{M_{H^\pm}}
\newcommand{\ssi}{\sigma_{SI}}
\newcommand{\pb}{\mathrm{pb}}
\newcommand{\ratio}{\ssi(\text{1-loop})/\ssi(\text{tree-level})}
\newcommand{\ssit}{\ssi(\text{tree-level})}
\newcommand{\ssil}{\ssi(\text{1-loop})}

\title{\bf Electroweak corrections to the direct detection cross section of inert higgs dark matter}
\author{
Michael Klasen\footnote{michael.klasen@uni-muenster.de}  ~and 
Carlos E. Yaguna\footnote{carlos.yaguna@uni-muenster.de} \\ 
\it \small Institut f\"ur Theoretische Physik, Universit\"at M\"unster,\\
\it \small Wilhelm-Klemm-Stra\ss e 9, D-48149 M\"unster, Germany\\[4mm]
Jos\'e D. Ruiz-\'Alvarez\footnote{Now at IPNL, Lyon, France. jose@gfif.udea.edu.co}\\ \it \small Instituto de F\'{i}sica, Universidad de Antioquia,\\
\it \small  A.A. 1226, Medell\'{i}n, Colombia}
\date{}

\begin{document}
\maketitle

\vspace*{-10cm}
\begin{flushright}
\texttt{MS-TP-13-01}
\end{flushright}
\vspace*{9cm}

\begin{abstract}
The inert higgs model is a minimal extension of the Standard Model that features a viable  dark matter candidate, the so-called inert higgs ($H^0$). In this paper, we compute  and analyze the dominant electroweak corrections to the direct detection cross section of dark matter within this model. These corrections arise from one-loop diagrams mediated by gauge bosons that, contrary to the tree-level result, do not depend on the unknown scalar coupling $\lambda$. We study in detail these contributions and show that they can modify in a significant way the prediction of the spin-independent direct detection cross section. In both viable regimes of the model, $\mh<M_W$ and $\mh\gtrsim 500~\gev$, we find regions where the cross section at one-loop  is  much larger than at tree-level. We also demonstrate that, over the entire viable parameter space of this model, these  new contributions bring the spin-independent   cross section within the reach of future direct detection experiments.
\end{abstract}

\section{Introduction}
Direct detection is possibly the most promising way of observing and identifying the dark matter --that mysterious form of matter that  accounts for about $20\%$ of the energy density of the Universe \cite{Komatsu:2010fb}. Direct detection experiments try to observe, via recoil-energy,  the scattering  of dark matter particles with nuclei and to determine from it some fundamental properties of the dark matter particle, such as its mass and its interactions. In recent years, these experiments, particularly XENON100 \cite{Aprile:2011dd,Aprile:2012nq},  have made outstanding progress in this regard and have started to exclude interesting regions of the parameter space of common models of dark matter --see e.g. \cite{Farina:2011bh, Strege:2011pk,Mambrini:2011ik,Melfo:2011ie}. In the near future, planned experiments, such as XENON-1T, will either find direct evidence of dark matter or increase the excluded regions even further. In both cases, it is of crucial importance to have reliable predictions for the direct detection cross section of dark matter. Otherwise, we would not be able to assess the implications of forthcoming data for specific models of dark matter or to foresee the extent to which  future facilities would be able to constrain them.

In most dark matter models, the one-loop electroweak contributions to the  dark matter direct detection cross section are expected to give only a tiny correction to the tree-level result, so there is no need, at least at present, to compute them. It may happen, however, that the tree-level result  features a strong suppression not necessarily present at higher orders. If that is the case, the calculation of such electroweak corrections becomes necessary if one wants to correctly predict the direct detection cross section of dark matter. It turns out that this situation actually arises in one of the most economical models that have been proposed to explain the dark matter puzzle: the inert doublet model \cite{Barbieri:2006dq,Ma:2006km,LopezHonorez:2006gr}.

In the inert doublet model, the Standard Model is extended with a second higgs doublet that is odd under a new $Z_2$ symmetry. The lightest component of this doublet becomes automatically stable and, if neutral, a good dark matter candidate, the so-called inert higgs ($H^0$). In recent years, the phenomenology of this model has been extensively studied in a number of works --see e.g. \cite{Gustafsson:2007pc,Lundstrom:2008ai,Agrawal:2008xz,Andreas:2009hj,Nezri:2009jd,Arina:2009um,Dolle:2009ft,Miao:2010rg, Honorez:2010re,LopezHonorez:2010tb,Grzadkowski:2010au,Martinez:2011ua,Borah:2012pu,Gustafsson:2012aj,Arhrib:2012ia}. In the inert doublet model, the tree-level direct detection cross section is determined by a higgs ($h$) mediated diagram and will  be suppressed whenever the coupling $H^0H^0h$, which is proportional to a free parameter of this model, becomes small. Since at one-loop, $H^0q$ scattering may proceed entirely via gauge processes ($W^\pm$ and $Z^0$ mediated diagrams), it is not guaranteed that these one-loop corrections will be smaller than the tree-level result. Motivated by this simple observation,  we calculate and analyze, in this paper, the dominant electroweak corrections to the  direct detection cross section of inert higgs dark matter. We will see that they may modify in a significant way the tree-level prediction within important regions of the viable parameter space, sometimes giving  the dominant contribution to the spin-independent direct detection cross section. Moreover, they always  bring this cross section within the reach of future experiments such as XENON-1T.

The rest of the paper is organized as follows. In the next section the inert doublet model of dark matter is briefly reviewed, outlining its parameter space and its viable regions.  Then, in section \ref{sec:1loop}, we present the calculation of the dominant electroweak corrections to the direct detection cross section of inert higgs dark matter and show its behavior  as a function of the parameters of the model. Sections \ref{sec:reslow} and \ref{sec:reslarge} contain our main results. They demonstrate the impact of these electroweak corrections within the two viable regimes of the model: the low mass one ($\mh<M_W$) in section \ref{sec:reslow} and  the large mass one ($\mh\gtrsim 500~\gev$) in section \ref{sec:reslarge}.  In both cases, we identify the regions where the corrections are expected to be important. To further substantiate our findings,  we perform a scan over the entire parameter space of the model and we analyze it in some detail. Finally, our conclusions are presented in section \ref{sec:con}.

\section{The inert doublet model}
\label{sec:idm}
The inert doublet model is a simple extension of the Standard Model
with one additional higgs doublet $H_2$ and an unbroken $Z_2$ symmetry,
under which $H_2$ is odd while all other fields are even. This
discrete symmetry prevents the direct coupling of $H_2$ to fermions and,
crucial for dark matter, guarantees the stability of the lightest
inert particle. The scalar potential of this model is given by
\begin{align}
V=&\mu_1^2|H_1|^2+\mu_2^2|H_2^2|+\lambda_1|H_1|^4+\lambda_2|H_2|^4+\lambda_3|H_1|^2|H_2|^2\nonumber\\
&+\lambda_4|H_1^\dagger H_2|^2+\frac{\lambda_5}{2}\left[(H_1^\dagger H_2)^2+\mathrm{h.c.}\right]\,,
\end{align}
where $H_1$ is the Standard Model higgs doublet, and $\lambda_i$ and
$\mu_i^2$ are real parameters. Four new physical states are obtained in
this model: two charged states, $H^\pm$, and two neutral ones, $H^0$
and $A^0$. Either of them could account for the  dark matter. In the
following, we assume that $H^0$ is the lightest inert particle,
$\mh^2<\ma^2,\mhc^2$, and, consequently,  the dark
matter candidate. After electroweak symmetry breaking, the inert scalar
masses take  the following form
\begin{align}
\mhc^2&= \mu_2^2+\frac12\lambda_3v^2 ,\nonumber \\
\mh^2&= \mu_2^2+\frac12(\lambda_3+\lambda_4+\lambda_5)v^2 ,\nonumber \\
\ma^2&= \mu_2^2+\frac12(\lambda_3+\lambda_4-\lambda_5)v^2\,, 
\end{align}
where $v=246$ GeV is the vacuum expectation value of $H_1$. Let us introduce at this point  the parameter $\lambda$ defined by
\begin{equation}
\lambda\equiv(\lambda_3+\lambda_4+\lambda_5)/2.
\end{equation}
This parameter is of particular relevance to our direct detection study as it determines the coupling $H^0H^0h$, and therefore the tree-level direct detection cross section --see next section.  In addition to $\lambda$,  it is convenient to take $\mh$, $\ma$, and $\mhc$ as the remaining free parameters of the inert sector. The tree-level direct detection cross section depends also on the  higgs  mass ($\mhiggs$). Given the small range to which  $\mhiggs$ has been  constrained by recent data \cite{:2012gk,:2012gu},  we have simply set $\mhiggs=125~\gev$ throughout this paper.

The new parameters of the inert doublet model are not entirely free, they are subject to a number of  theoretical and experimental constraints  --see  e.g. \cite{Barbieri:2006dq} and
\cite{LopezHonorez:2006gr}. The requirement of vacuum stability imposes that
\begin{equation}
\lambda_1,\lambda_2 > 0\,,\qquad 
\lambda_3, \lambda_3+\lambda_4-|\lambda_5|>-2\sqrt{\lambda_1\lambda_2}\,.
\end{equation}
LEP data constrain the mass of the charged scalar, $\mhc$,  to be larger than about $90\,\gev$ \cite{Pierce:2007ut} while  some regions in the plane ($\mh,\ma$) are also excluded,  see \cite{Lundstrom:2008ai}. In addition, the inert doublet, $H_2$,  contributes to electroweak precision parameters such as $S$ and $T$, which must be small to remain compatible with current data. Finally,  the relic density of inert higgs dark matter should be compatible with the observed dark matter density \cite{Komatsu:2010fb}. To evaluate $\oh$, we have used micrOMEGAs \cite{Belanger:2010gh}, which automatically takes into account resonances and coannihilation effects. Into  micrOMEGAs we have incorporated the annihilation into the three-body final state $WW^*$ ($H^0H^0\to WW^*\to W f\bar f'$) which modifies in a significant way the predicted relic density for $\mh\lesssim M_W$ \cite{Honorez:2010re}.

In previous works \cite{LopezHonorez:2006gr}, it had been found that the dark matter constraint
can not  be satisfied  for arbitrary values of $\mh$.  Two
separate regions remain viable\footnote{Notice that, as anticipated in \cite{LopezHonorez:2010tb}, the \emph{new viable region}, $M_W<\mh\lesssim 150~\gev$,  has already been excluded by the recent XENON100 data \cite{Aprile:2011dd,Aprile:2012nq}.}, one at low masses and the other at large masses. In the low mass regime ($\mh\lesssim M_W$), the annihilation of dark matter is dominated by either the $b\bar b$ final state or the three-body final state $WW^*$, and may be enhanced due to the presence of the higgs resonance at $\mh\sim \mhiggs/2$. Moreover, $H^0$-$A^0$ coannihilations  may also play a role in the determination of the dark matter relic density. In the large mass regime ($\mh>500~\gev$), dark matter annihilates either into gauge bosons ($W^+W^-$, $Z^0Z^0$) or into higgses. These annihilation channels are usually  very efficient, so the relic density tends to be suppressed. The observed value of the dark matter density  can still be obtained in this regime but only when the mass splitting between the inert particles is tiny. Since these two dark matter compatible regimes have completely different phenomenologies, we will split our analysis and discuss our main results in two different  sections, one dedicated to each regime.  Before that, we present, in the next section, the calculation of the electroweak corrections to the spin-independent cross section and  obtain some  preliminary results.
\section{The direct detection cross section at one-loop}
\label{sec:1loop}
\begin{figure}[t]
\begin{center}
\includegraphics[scale=0.4]{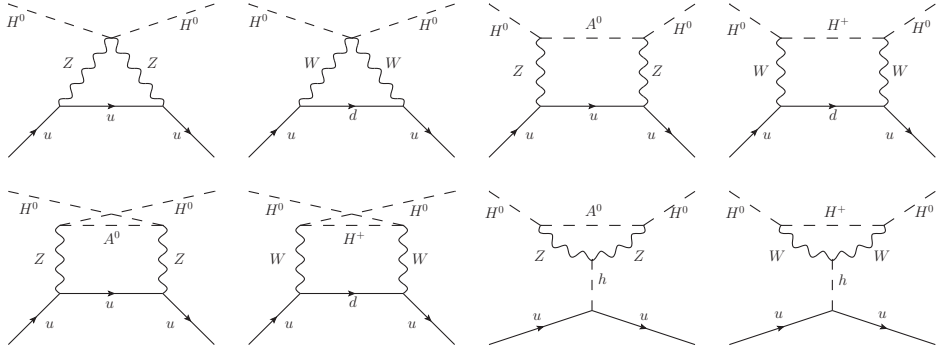}
\caption{\small \it  The Feynman diagrams that give the dominant corrections to the direct detection cross section of inert higgs dark matter. 
\label{fig:diags}}
\end{center}
\end{figure}

\begin{figure}[t]
\begin{center}
\includegraphics[scale=0.4]{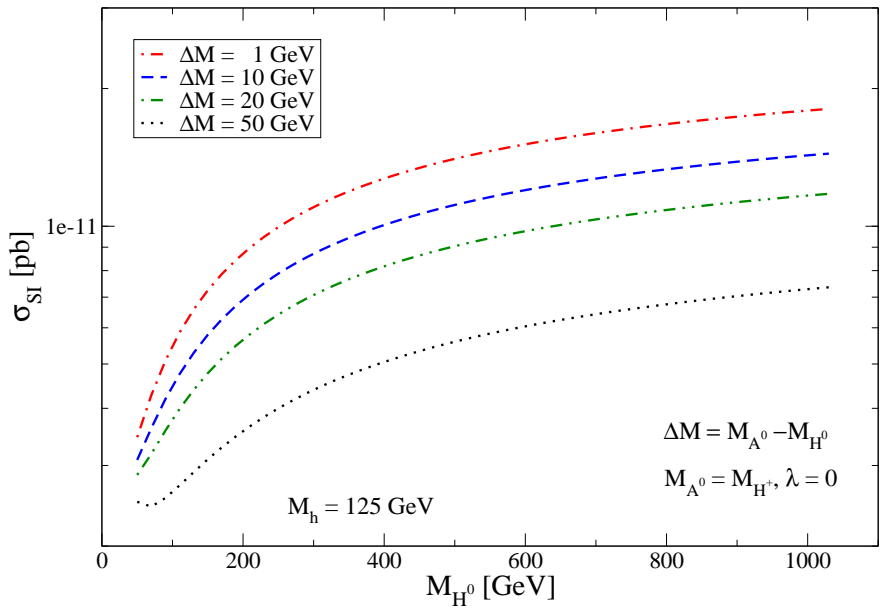}
\caption{\small \it  The purely one-loop contribution to the spin-independent cross section as a function of the dark matter mass for four different values of $\Delta M=\ma-\mh=\mhc-\mh$: $1$, $10$, $20$, and $50$ GeV (from top to bottom). In this figure we have set $\lambda=0$ and $\mhiggs=125~\gev$. 
\label{fig:simh0}}
\end{center}
\end{figure}

In the inert doublet model, the dark matter direct detection cross section at tree-level is given by 
\begin{equation}
 \ssit=\frac{m_r^2}{\pi}\left(\frac{\lambda}{\mh\mhiggs^2}\right)^2 f^2 m_N^2,
\label{eq:ddtree}
\end{equation}
where $f$ is a quark form factor\footnote{In our numerical evaluations, we use for the quark form factors $f_q$ the default values from micrOMEGAs \cite{Belanger:2010gh}.} and $m_r$ is the reduced mass of the dark matter-nucleon system. This cross section arises from a higgs mediated diagram and is seen to be proportional to $\lambda^2$. In many models, the tree-level value of $\sigma_{SI}$ is accurate enough for most purposes and there is no need to compute  electroweak corrections to it. The inert doublet model, however, may be an exception to that rule. In fact, in this model not only can the coupling $\lambda$  be very small  (much smaller than the gauge couplings), but there are one-loop diagrams mediated by the gauge bosons that contribute to $\ssi$  which  do not depend on $\lambda$ and are instead entirely determined by the gauge couplings and the masses of the inert particles. It is quite possible, therefore, that the tree-level result, equation (\ref{eq:ddtree}), fails to give the correct prediction for the spin-independent direct detection cross section in certain regions of the parameter space. For that reason, in this paper we compute the dominant electroweak corrections to $\ssi$ and we analyze their importance in both the low and the large mass regime of the model. A calculation similar to this was first presented in \cite{Cirelli:2005uq} and later applied to the inert doublet model in \cite{Hambye:2009pw}. It must be emphasized, however, that the model  in \cite{Cirelli:2005uq} is not exactly the inert higgs model and that they considered only the regime $M_{DM}\gg M_W$. Since their results can not be directly used for our study,  we have calculated these corrections ourselves without making any assumptions on the masses of the inert particles.  We limit ourselves to  those diagrams which might become dominant when $\lambda$ is small, that is to diagrams mediated by electroweak gauge bosons and independent on $\lambda$. 
The contributing diagrams are shown in figure \ref{fig:diags}. In the following, we denote by $\ssit$ or simply by $\ssi$ the value of the spin-independent cross section that is obtained when these diagrams are taken into account.  Notice that these electroweak corrections  depend only on three unknowns\footnote{The total  amplitude (tree + one-loop) will depend also on $\lambda$ and $\mhiggs$. Since the latter is fixed, the total amplitude depends on $4$ parameters.}: $\mh$, $\ma$, and $\mhc$. Next, we will numerically study $\ssi$ as a function of these parameters and we will demonstrate that these one-loop contributions may indeed  be larger than the tree-level result.

Figure \ref{fig:simh0} shows the purely one-loop contribution ($\lambda=0$) to $\ssi$ as a function of $\mh$ for four different values of the mass splitting $\Delta M=\ma-\mh=\mhc-\mh$. From top to bottom, the lines correspond to $\Delta M=1,10,20,50~\gev$. Notice that the electroweak corrections  give a cross section  of order $10^{-11}~\pb$ - $10^{-12}~\pb$   depending slightly on the dark matter mass and on the mass splitting.   $\ssi$ initially increases with $\mh$ but then reaches a constant value for  large  $\mh$ --a result compatible with that found in \cite{Cirelli:2005uq}. It is also clear from the figure that $\ssi$ decreases with the mass splitting between the inert particles. 

\begin{figure}[tb]
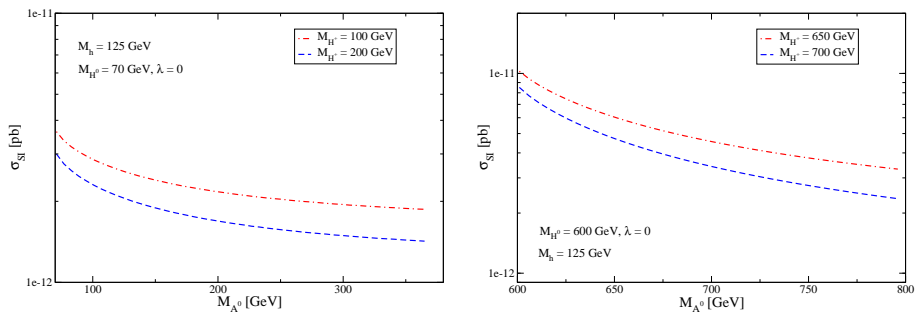

\begin{center}
\begin{tabular}{cc}
\includegraphics[scale=0.2]{silight.eps} & \includegraphics[scale=0.2]{siheavy.eps}
\end{tabular}
\caption{\small \it  The purely one-loop contribution to the spin-independent cross section as a function of $\ma$ for different sets of parameters. In the left panel, we consider a light dark matter particle, $\mh=70~\gev$, and $\mhc=100,200~\gev$. In the right panel, a heavy dark matter candidate is considered, $\mh=600~\gev$, and $\mhc=650,700~\gev$. In both panels, $\lambda=0$ and $\mhiggs=125~\gev$.
\label{fig:silightheavy}}
\end{center}
\end{figure}

It is also interesting to look at the behavior of $\ssi$ as a function of $\ma$ (or $\mhc$) for a fixed value of the dark matter mass. In figure \ref{fig:silightheavy} we illustrate that for the low mass regime ($\mh=70~\gev$, left panel) and the heavy mass regime ($\mh=600~\gev$, right panel). In each panel two different values of $\mhc$ are considered.  In both regimes  we find that $\ssi$ decreases with $\ma$ and with $\mhc$ and that it varies between  $10^{-11}~\pb$ and $10^{-12}~\pb$, as found before.  

\begin{figure}[tb!]
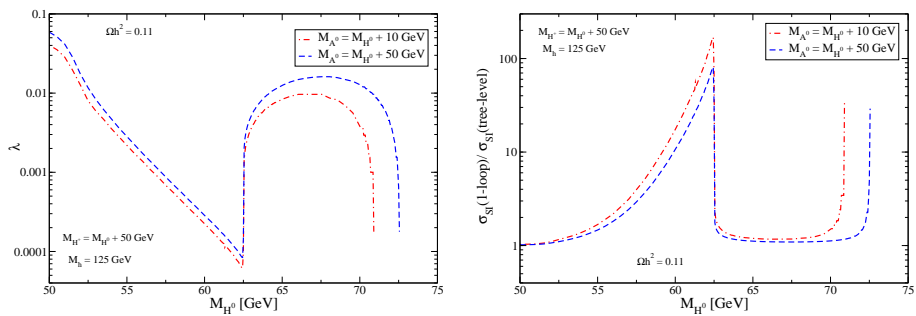

\begin{center}
\begin{tabular}{cc}
\includegraphics[scale=0.2]{pslight.eps} & \includegraphics[scale=0.2]{sicorrlight.eps} 
\end{tabular}
\caption{\small \it  Left: The viable parameter space of the inert doublet model in the plane ($\mh,\lambda$) for two different values of $\ma-\mh$: $10~\gev$ (dotted-dashed line) and $50~\gev$ (dashed line).  In this figure, $\mhc-\mh$ was set to $50~\gev$ and $\mhiggs$ to $125~\gev$. Along the lines, the dark matter constraint, $\oh=0.11$, is satisfied. Notice that the coupling $\lambda$ can reach values as  small as $10^{-4}$. Right: The correction to the spin-independent direct detection cross section as a function of the dark matter mass along the viable regions from the left panel.
\label{fig:pslight}}
\end{center}
\end{figure}

So far in our analysis we have made two important simplifications: \emph{i)} we have set $\lambda=0$, or equivalently we have limited ourselves to the purely one-loop contribution; \emph{ii)} we have not yet enforced the constraints on the parameters of the inert doublet model. In the next two sections, where our main results are presented, we will get rid of these simplifications. Ultimately,  what we actually want to know is how important these electroweak corrections are within the viable regions of the inert doublet model. In particular, we would like to determine if they can give the dominant contribution to $\ssi$ and in which regions that happens. We also want to know how these corrections modify the prospects for the direct detection of dark matter in future experiments. To that end, we should move away from the $\lambda=0$ limit considered in this section and we should ensure that $\ssi$ is evaluated only for models that are compatible with all the known phenomenological and cosmological constraints. 

\section{Results for the low  mass regime}
\label{sec:reslow}
In this section, we examine the implications of the electroweak corrections to $\ssi$ within the low mass regime of the inert doublet model. To begin with, we show, in the left panel of figure \ref{fig:pslight},  the viable parameter space in the plane ($\mh,\lambda$) for $\mhc=\mh+50~\gev$ and two different values of $\ma-\mh$: $10~\gev$ and $50~\gev$. Along the lines, the dark matter relic density is compatible with current observations, $\oh=0.11$.  Since coannihilation effects are important for $\ma=\mh+10~\gev$ (dash-dotted line) the required value of $\lambda$ is always smaller than that for $\ma=\mh+50~\gev$ (dashed line), where they are not.  Close to the higgs resonance, $\mh=\mhiggs/2=62.5~\gev$, the annihilation of dark matter tends to be very efficient, so $\lambda$ has to be very small to avoid depleting the abundance of dark matter in the early Universe. For $\mh\sim 70-72~\gev$, the annihilation into the three-body final state $WW^*$ \cite{Honorez:2010re}, a process dominated by the gauge interactions, is  sufficient to account for the observed dark matter so $\lambda$ must be small to suppress the additional higgs-mediated annihilations (whose strength increases with $\lambda$).  The main lesson from this figure is  that  there are regions  in the viable parameter space of  the inert doublet model where the scalar coupling $\lambda$ is indeed much smaller than the gauge couplings, reaching values as low as $10^{-4}$.
\begin{figure}[tb!]
\begin{center}
 \includegraphics[scale=0.4]{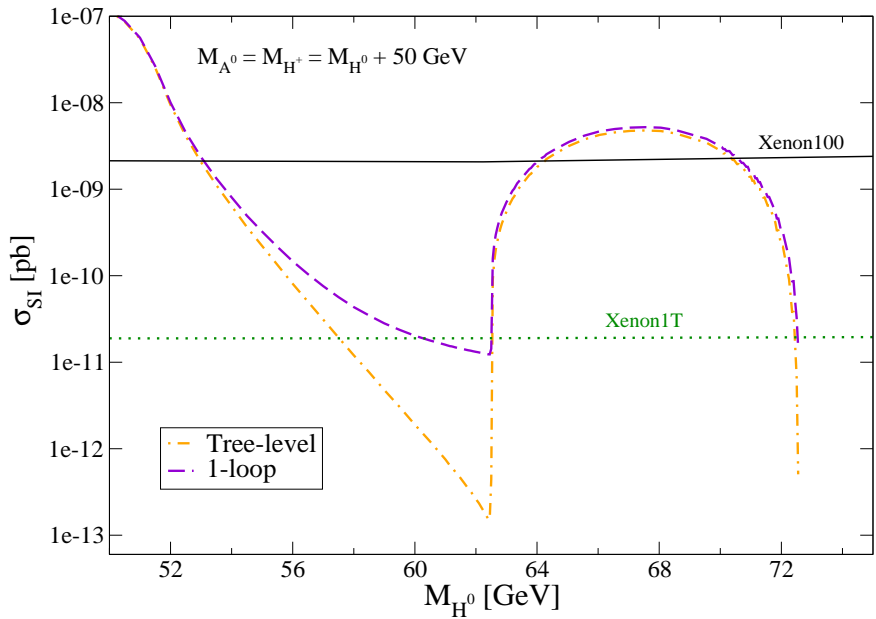}
\caption{\small \it  A comparison between the tree-level and the one-loop  direct detection cross section for one of the viable regions of figure \ref{fig:pslight}. 
\label{fig:sicorrlight}}
\end{center}
\end{figure}

In such regions, we expect the  one-loop corrections to modify in a significant way the prediction of the inert higgs direct detection cross section, and perhaps to give a contribution larger than the tree-level one. To illustrate the effect of the electroweak corrections, in the following we will either compare $\ssit$ with $\ssil$ in the same figure or study their ratio as a function of the parameters of the model. 

 The right panel of figure \ref{fig:pslight} shows the ratio $\ssil/\ssit$ along the viable lines from the left panel. As expected, the correction is large where $\lambda$ is small and vice versa. We see that the one-loop correction can indeed be much larger than the tree-level result, with $\ratio$ reaching values as high as $\sim 30$ for $\mh\sim 71~\gev$ and $\sim 100$ for $\mh\sim \mhiggs/2$. Outside these regions, the correction is small but not necessarily negligible and may easily account for a $20\%$ increase in $\ssi$.

\begin{figure}[tb!]
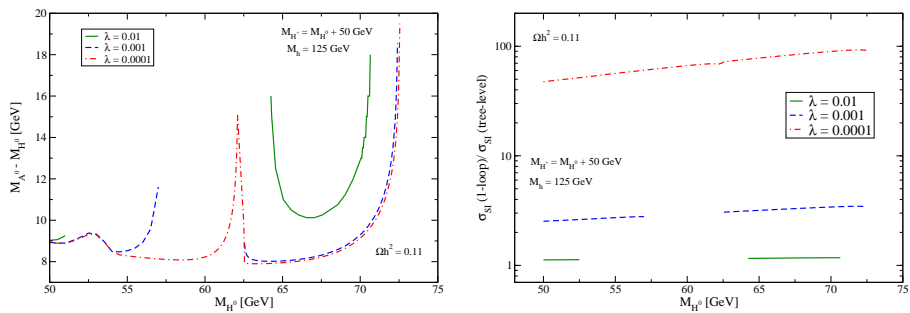

\begin{center}
\begin{tabular}{cc}
\includegraphics[scale=0.2]{coannps.eps} &\includegraphics[scale=0.2]{sicorrcoann.eps}
\end{tabular}
\caption{\small \it  Left: The parameter space of the inert doublet model in the plane ($\mh,\ma-\mh$)  for three different values of the coupling $\lambda$: $10^{-2}$ (solid line),  $10^{-3}$ (dashed line) and $10^{-4}$ (dashed-dotted line).  The value of $\mhc$ was set to $\mh+50~\gev$. Along the lines the relic density constraint, $\oh=0.11$, is satisfied --mainly via $H^0$-$A^0$ coannihilations. Notice that for $\lambda=10^{-2}$ and  $\lambda=10^{-3}$ there is a range in $\mh$ with no viable points. Right: The correction to the spin-independent direct detection cross section as a function of the dark matter mass along the viable lines from the left panel. Notice that in this case the correction only slightly depends on $\mh$.
}
\label{fig:pscoann}
\end{center}
\end{figure}

A direct comparison between $\ssil$ and $\ssit$ is shown  in figure \ref{fig:sicorrlight}. Here, we have selected, from the two viable lines discussed in the previous figure, the one featuring $\ma=\mh+50~\gev$.  For illustration, the current bound from XENON100 and the expected sensitivity of XENON-1T are also displayed. The former already excludes the regions $\mh>53~\gev$ and $64<\mh/\gev<70$ in this parameter space. The effect of the one-loop corrections is clearly seen close to the higgs resonance, where it prevents the cross section from going below about $10^{-11}~\pb$.  A similar effect takes place also at the largest allowed value of $\mh$. 

One may be tempted to conclude, from  the above figures, that in the low mass regime the one-loop corrections to $\ssi$ can become very large  only around two specific values of $\mh$, $\mhiggs/2$ and $71~\gev$, and that they are much smaller  everywhere else. That such a conclusion is wrong --is only an artifact of the specific slice of the parameter space being displayed-- is demonstrated by figure \ref{fig:pscoann}. Its left panel shows the viable regions for $\lambda=10^{-2},10^{-3},10^{-4}$ and $\mhc=\mh+50~\gev$. Because in this case $H^0$-$A^0$ coannihilations play a prominent role in obtaining the right value of the dark matter density, it makes sense to display the parameter space in the plane ($\mh$, $\ma-\mh$). For $\lambda=10^{-4}$ (dash-dotted line) it is always possible to find a value of $\ma-\mh$ that gives the observed value of the dark matter density, but that is not true for $\lambda=10^{-3}$ or $\lambda=10^{-2}$. Notice that the required mass splitting  increases significantly close to the higgs resonance and near the maximum allowed value of $\mh$. The small bump observed at $\mh\sim 52.5~\gev$ is due to the effect of resonant $A^0$-$A^0$ annihilations on the relic density.  The right panel of figure  \ref{fig:pscoann} shows the  correction to $\ssi$ along such viable regions. We see that in this case the correction does not strongly depend on $\mh$: it is of order of several percent for $\lambda=10^{-2}$ (solid line),  a factor 2 to 4 for  $\lambda=10^{-3}$ (dashed line) and it reaches almost a factor 100 for  $\lambda=10^{-4}$ (dash-dotted line). In all cases there is a slight increase in the correction with the dark matter mass. Clearly, large  electroweak corrections to $\ssi$  are not  confined to $\mh\sim\mhiggs/2$ and $\mh\sim 70~\gev$  but can actually be found for any value of $\mh$. At the end, it is the size of $\lambda$ and not $\mh$ what determines how important the corrections are, and $\lambda$  can vary over several orders of magnitude within the viable regions of the model.    

\begin{figure}[tb]
\begin{center}
\includegraphics[scale=0.35]{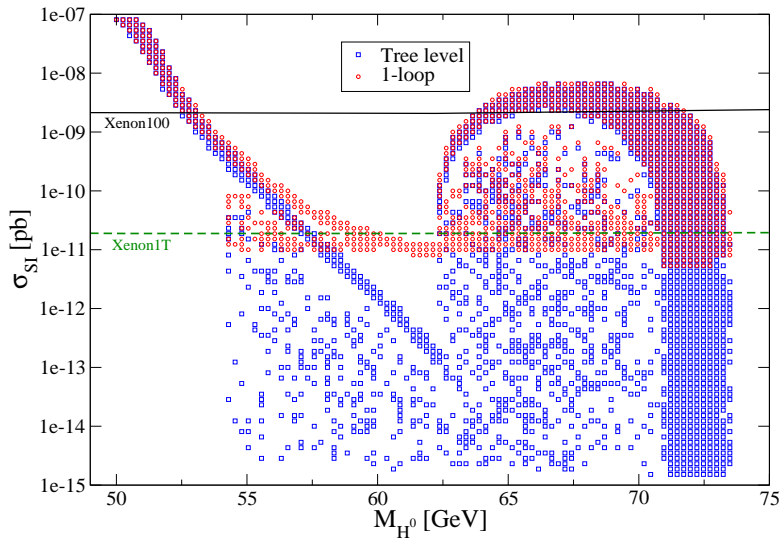}
\caption{\small \it  A scatter plot of the  spin-independent direct detection cross section at tree-level and at one-loop as a function of the dark matter mass. In this figure all the parameters of the inert higgs model were allowed to vary randomly (see text for details) and all experimental bounds were taken into account. 
\label{fig:scansilight}}
\end{center}
\end{figure}

To assess in all generality, and independently of the specific slice of parameter space examined,  the relevance of the electroweak corrections to $\ssi$, we have scanned, using Markov Chain Monte Carlo techniques \cite{Baltz:2004aw}, the entire parameter space of the inert doublet model. After allowing the parameters to vary within the following ranges
\begin{align}
 80 ~\gev>\mh&> 50~\gev,\\
 \ma&>\mh,\\
\mhc&>90~\gev,\\
1>\lambda&>10^{-5},
\end{align}
and imposing all the experimental bounds (collider, precision, dark matter, etc.) we obtained a sample of about $10^4$ viable models to analyze. Figure \ref{fig:scansilight} shows a scatter plot of these models in the plane ($\mh, \ssi$). The (blue) squares show $\ssit$ and the (red) circles $\ssil$. Two classes of models can be easily distinguished in this figure: the \emph{annihilating} models that are concentrated along a narrow band similar to that observed in figure \ref{fig:sicorrlight} and the \emph{coannihilating} models which are scattered in the region below that band. They are absent below $\mh\sim 55~\gev$ because the  mass splitting required for coannihilations to be important becomes inconsistent with collider bounds \cite{Lundstrom:2008ai}. Notice that whereas $\ssit$ may be as small as $10^{-15}~\pb$, $\ssil$ does not go below $10^{-11}~\pb$ or so. From the figure we also see that some regions are already excluded by the XENON100 bound \cite{Aprile:2012nq} (solid line). The most important result, however, is the fact that the one-loop corrections always bring $\ssi$ within the reach of future direct detection experiments and in particular very close to the XENON-1T expected sensitivity.

\begin{figure}[tb]
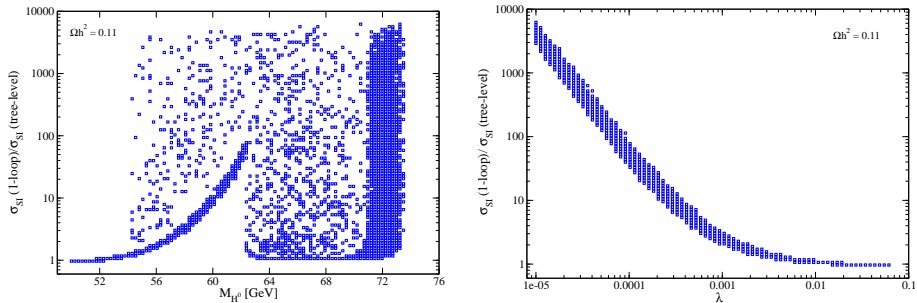

\begin{center}
\begin{tabular}{cc}
\includegraphics[scale=0.2]{scancomplight.eps} & \includegraphics[scale=0.2]{scancorrlight.eps}
\end{tabular}
\caption{\small \it Left: A scatter plot of   $\ratio$ as a function of $\mh$.  Right:  A scatter plot of $\ratio$ as a function of $\lambda$. In this figure all the parameters of the inert higgs model were allowed to vary randomly (see text for details) and all experimental bounds were taken into account. 
\label{fig:scanlight}}
\end{center}
\end{figure}
Figure \ref{fig:scanlight} shows the same sample of viable models, but in two additional planes. The right panel shows $\ratio$  as a function of $\mh$. The annihilating and coannihilating models can again be clearly distinguished in this figure. Notice that the correction can be very large, say $\ratio\sim 100$, pretty much for any value of $\mh$. The right panel displays  the same ratio but now as a function of $\lambda$. The general behavior is as anticipated, with the correction increasing for decreasing $\lambda$. It can also be seen in  this figure that $\ssil$ becomes larger than $\ssit$ for $\lambda\gtrsim 10^{-3}$, as we had found before. The small spread observed in this figure clearly demonstrates that it is the size of $\lambda$ that determines how large $\ratio$ is.

Summarizing, we have seen that in the small mass regime of the inert doublet model, $\mh<M_W$, the electroweak corrections to the spin-independent direct detection cross section can be quite relevant, giving in certain cases the dominant contribution to $\ssi$. We have observed that these  corrections become large when $\lambda\lesssim 10^{-3}$. Such values of $\lambda$ are  compatible with the dark matter constraint thanks to coannihilations (for a wide range of $\mh$), resonant annihilations (for $\mh\lesssim \mhiggs/2$), or annihilations into three-body final states (for $\mh\sim 72~\gev$). We have also noticed that in contrast to $\ssit$, which can be arbitrarily small, $\ssil$ is never below $\sim10^{-11}~\pb$. Thus, over the entire low mass regime, the electroweak corrections we have studied bring $\ssi$ within the reach of future direct detection experiments.

\section{Results for the large mass regime}
\label{sec:reslarge}
\begin{figure}[tb]
\begin{center}
\includegraphics[scale=0.4]{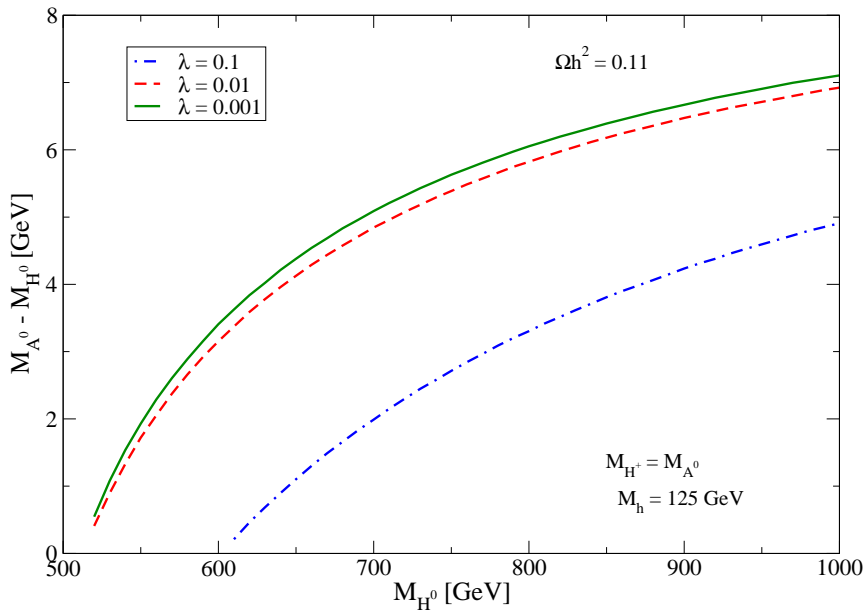}
\caption{\small \it  The viable parameter space of the inert doublet model in the plane ($\mh,\ma-\mh$) for different values of $\lambda$. In this figure we consider the large mass regime of the model and we set $\mhc=\ma$ and $\mhiggs=125~\gev$. Notice that the required mass splitting is always very small. 
\label{fig:psheavy}}
\end{center}
\end{figure}
\begin{figure}[tb]
\begin{center}
\includegraphics[scale=0.4]{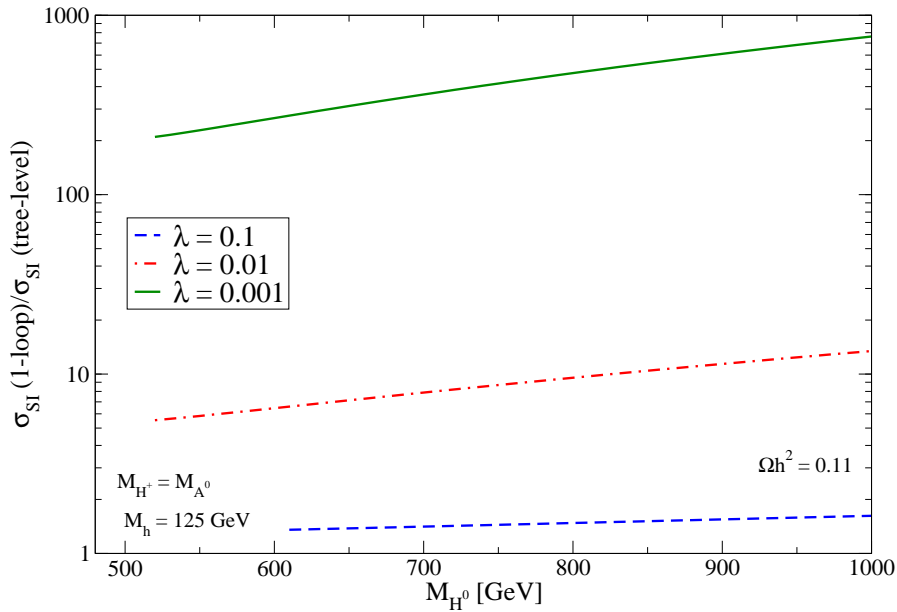}
\caption{\small \it  The  correction to the spin-independent direct detection cross section as a function of the dark matter mass along the viable lines of figure \ref{fig:psheavy}.
\label{fig:sicorrheavy}}
\end{center}
\end{figure}

We now focus our attention on the heavy mass regime of the model, $\mh\gtrsim 500~\gev$. 
Figure \ref{fig:psheavy} shows viable regions of the inert doublet model in the plane ($\mh$, $\ma-\mh$) for different values of the scalar coupling $\lambda$. For concreteness, in this figure we have set $\mhc=\ma$ and we have restricted the mass range to $\mh<1~\tev$. Notice that even though the mass splitting between the inert particles increases with the dark matter mass, it is always very small (below the per cent level). At $\mh=1~\tev$, for instance, it amounts to no more than $7~\gev$. This is a generic and well-known feature of the  large mass regime of the inert doublet model: only for small values of $\ma-\mh$ and $\mhc-\mh$ can the relic density constrained be satisfied, see e.g. \cite{Hambye:2009pw}. In the figure we see that the viable parameter space starts at $\mh\sim 520~\gev$ for $\lambda=10^{-2},10^{-3}$ and around $600~\gev$ for $\lambda=0.1$. In this regime there are neither resonances nor thresholds, so the analysis is much simpler. As we saw in figure \ref{fig:simh0}, the one-loop correction to $\ssi$ initially increases with $\mh$ whereas the tree-level value of $\ssi$ decreases with $\mh^2$ (see equation \ref{eq:ddtree}). Since, in addition, $\lambda$ can be made arbitrarily small in this regime, we expect that the electroweak corrections to $\ssi$ be more relevant than for the low mass regime. Figure \ref{fig:sicorrheavy} shows $\ratio$ along the viable lines of figure \ref{fig:psheavy}.  As expected, the correction is larger the smaller $\lambda$ is. We also observe  that as $\mh$ increases,  the correction indeed becomes more important. It amounts to a factor between $1$ and $2$  for $\lambda=0.1$, about a factor $10$ for $\lambda=0.01$, and more than  $200$ for $\lambda=10^{-3}$. Notice, for example, that in the large mass regime $\ratio\sim 2$ can be obtained already for $\lambda=0.1$ whereas in the low mass regime that would require a value of $\lambda$ at least one order of magnitude smaller.

\begin{figure}[tb]
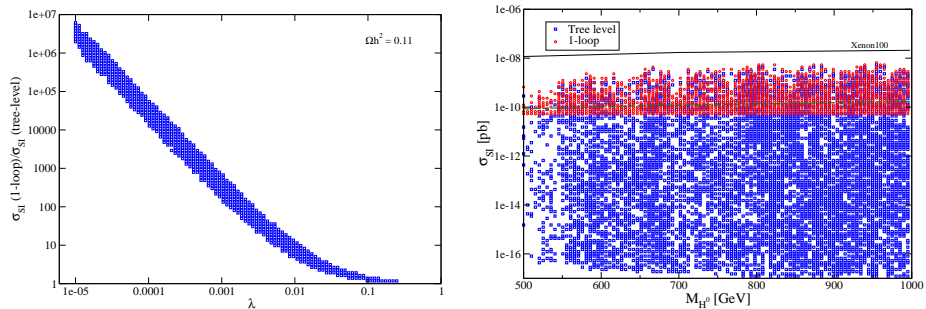

\begin{center}
\begin{tabular}{cc}
\includegraphics[scale=0.2]{scancorrheavy.eps} & \includegraphics[scale=0.2]{scansiheavy.eps}
\end{tabular}
\caption{\small \it Some results of the scan over the parameters of the model for the large mass regime. Left: A scatter plot of the correction to $\ssi$ as a function of $\lambda$. Right: A comparison between the tree-level and  the one-loop value of $\ssi$ as a function of $\mh$. The solid line shows the current bound from XENON100 whereas the dashed line corresponds to the expected sensitivity of XENON-1T.
\label{fig:scanheavy}}
\end{center}
\end{figure}

We have also scanned the parameter space of this regime by allowing the inert masses to vary in the range
\begin{align}
 1~\tev>\mh&>500~\gev,\\
\mhc&>\mh,\\
\ma&>\mh.
\end{align}
After imposing all the relevant constraints, we obtained a sample of approximately $10^4$ viable models. Figure \ref{fig:scanheavy} shows this sample of models in two different planes. The left panel displays $\ratio$ as a function of $\lambda$. It demonstrates that $\ratio$ is a decreasing function of $\lambda$, as expected, and that it becomes much larger than $1$, say $\sim 10$, for  $\lambda\sim 10^{-2}$. The small spread of models in this plane again indicates that it is fundamentally $\lambda$ the parameter that determines the size of $\ratio$. The right panel compares the tree-level and one-loop value of $\ssi$ as a function of $\mh$. Notice that whereas at tree-level $\ssi$ could be as small as $10^{-17}~\pb$ at one-loop it is never below $10^{-11}~\pb$. From the figure we see that this region is not being currently probed by direct detection experiments --see  the present XENON100 bound (solid line). The future prospects, however, are very good because the one-loop corrections bring $\ssi$ within the reach of planned direct detection experiments. 

\section{Conclusions}
\label{sec:con}
We have computed and studied the dominant electroweak corrections to the direct detection cross section of inert higgs dark matter. These corrections arise from one-loop diagrams mediated by the electroweak gauge bosons and do not depend on the scalar coupling $\lambda$ that controls the tree-level cross section. We have analyzed the behavior of these one-loop contributions as a function of the parameters of the model, and have calculated their effect within the regions that are compatible with the dark matter constraint for the two distinct regimes of this model: the low mass regime ($\mh<M_W$) and the large mass regime ($\mh\gtrsim 500~\gev$). In both regimes, we have found regions where the one-loop corrections  not only become significant but can even be  larger than the tree-level result. In the low mass regime, this happens when $\lambda\lesssim 10^{-3}$, a value that can be compatible with the dark matter constraint via annihilation through the higgs resonance, annihilation into the three-body final state $WW^*$, or coannihilations. The first two require respectively $\mh\sim \mhiggs/2$ and $\mh\sim 72~\gev$ whereas coannihilations allow for a much wider  range of $\mh$. In the heavy mass regime, we found the effect of the electroweak corrections to be larger, with   corrections of order $100\%$ already for $\lambda=0.1$.  Thus, they must be necessarily taken into account when assessing the prospects for the direct detection of inert higgs dark matter.  From the scans over the full parameter space of the model, we also observed that these one-loop contributions always bring $\ssi$ within the reach of future direct detection experiments. 
\section*{Acknowledgments}
This work is supported by the ``Helmholtz Alliance for Astroparticle Phyics HAP''
 funded by the Initiative and Networking Fund of the Helmholtz Association. J.D.R. would like to thank D. Restrepo for his collaboration. 

\bibliographystyle{unsrt}
\bibliography{darkmatter}

\end{document}